\documentclass[doublecol]{epl2}
\usepackage{graphicx}
\usepackage{amssymb,amsmath,amsfonts,palatino,amsthm}
\usepackage{amssymb}
\usepackage{epstopdf}
\usepackage{color}

\title{From Causal Dynamical Triangulations \\ To Astronomical Observations}

\author{Jakub Mielczarek}

\institute{Institute of Physics, Jagiellonian University, {\L}ojasiewicza 11, 30-348 Cracow, Poland}


\abstract{
This letter discusses phenomenological aspects of dimensional reduction 
predicted by the Causal Dynamical Triangulations (CDT) approach to quantum 
gravity. The deformed form of the dispersion relation for the fields defined on 
the CDT space-time is reconstructed. Using the \emph{Fermi} satellite observations 
of the GRB 090510 source we find that the energy scale of the dimensional 
reduction is $E_* >   0.7 \sqrt{4-d_{\rm UV}}  \cdot 10^{10}$ GeV at (95 $\%$ CL),
where $d_{\rm UV}$ is the value of the spectral dimension in the UV limit.
By applying the deformed dispersion relation to the cosmological perturbations 
it is shown that, for a scenario when the primordial perturbations are formed in the 
UV region, the scalar power spectrum $\mathcal{P}_S \propto k^{n_S-1}$ 
where $n_S-1\approx \frac{3 r (d_{\rm UV}-2)}{(d_{\rm UV}-1)r-48}$. Here,   
$r$ is the tensor-to-scalar ratio. We find that within the considered model, 
the predicted from CDT deviation from the scale-invariance ($n_S=1$) is 
in contradiction with the up to date \emph{Planck} 
and \emph{BICEP2}. 
}

\begin{document}

\maketitle 

\section{Introduction}

The littleness of the Planck length is indescribable. The immense size the observable Universe 
is paralyzing. However, the geometric mean of the two characteristic values results in a tangible 
quantity. Explicitly, $\sqrt{ 1 l_{\rm Pl}\cdot 1 {\rm Gpc}}\approx 1 {\rm mm}$, where 
the Planck length  $l_{\rm Pl} \approx 1.62 \cdot 10^{-35}$ m  and the gigaparsec 
 $1 {\rm Gpc} \approx 3.09 \cdot 10^{28}$ m. 

As we will see from the analysis performed in the letter, better understanding of 
both the super-short and super-large scales can be gained thanks to this property. 
   
So, how does it work? There are two main possibilities. The first one is a result of 
accumulation of the tiny Planck scale deviations acting on particles propagating 
across cosmological distances. The second possibility exploits the evolution of the 
Universe.  The Universe is as big as it is because it is expanding. As a result, the 
Planck scale in the early universe, multiplied by the growth factor becomes 
macroscopic in the mature universe. In both cases, the multiplication law brings 
the Planck scale closer to the scales of our perception.  
   
We do not know yet what the Planck scale physics is. However, by taking the use of 
the compensation of scales discussed above, one may already try to falsify the theoretical 
models of quantum gravitational phenomena\footnote{Which are expected to occur
at the Planck scale.}, which we have. 

The theory of Planck scale physics which we are going to examine here are Causal 
Dynamical Triangulations (CDT) \cite{Ambjorn:2000dv,Ambjorn:2012jv}. The theory is a 
non-perturbative approach based on the path integral formulation of quantum mechanics. 
The crucial ingredient of the theory is a causality condition imposed at the level of Wick-rotated 
time coordinate. In similarity to statistical physics, the different space-time configurations are 
evaluated using Monte Carlo computer simulations. 

Various promising results arose from numerical studies of the quantum gravitational 
system with the positive cosmological constant. One of the most profound is the emergence
of four dimensional de Sitter space-time \cite{Ambjorn:2004qm,Ambjorn:2007jv} 
in the so-called phase C\footnote{Existence of the three different phases of CDT (A, B and C) has
been firmly confirmed so far \cite{Ambjorn:2012ij}.}. However, as reflected by features of the process 
of diffusion taking place at the simplicial manifold of CDT, the reconstructed de Sitter space-time is not 
fully classical.  

In particular, it has been shown that the spectral dimension\footnote{The precise definition 
will be given in the subsequent section.} characterizing the diffusion process can be parametrized 
in the following way \cite{Ambjorn:2005db}:  
\begin{equation}
d_S(\sigma) =  a -\frac{b}{c+\sigma},
\label{SpectralParametrization}
\end{equation}
where $\sigma$ is the diffusion time. At the representative point in phase C ($\kappa_0=2.2$, 
$\Delta=0.6$) the values $a=4.02$, $b=119$ and $c=54$ \cite{Ambjorn:2005db} and  
$a=4.06$, $b=135$ and $c=67$ \cite{Coumbe:2014noa} have been obtained from numerical 
simulations. In both cases, four dimensional space-time ($d_S \approx 4$) is correctly recovered
in the IR limit ($\sigma \rightarrow \infty$). However, in the UV limit ($\sigma \rightarrow 0$) dimensional
reduction to $d_S \approx 2$ was observed. This effect is not reserved to CDT only and has 
been noticed in other approaches to quantum gravity as well \cite{Carlip:2009kf}. 
Moreover, recent numerical studies revealed that the UV behavior of the spectral 
dimension may change depending on which part of the phase C is considered. In particular, 
in has been shown that when $C-A$ transition line is approached, the UV limit of 
the spectral dimension is consistent with  $d_S \approx 3/2$ \cite{Coumbe:2014noa}. 
Worth noticing is that, such value of the spectral dimension may have theoretical relevance 
(see Ref. \cite{Laiho:2011ya}). In what follows, we will refer to two different UV values
of the spectral dimension mentioned above. However, the performed calculations will 
allow to consider also other possible values of the spectral dimension in the UV limit.

The expression  (\ref{SpectralParametrization}) is a starting point for our further considerations. 
After fixing the IR value $d_S$ to be precisely equal to the topological space-time dimension 
$d=4$ and we obtain
\begin{equation}
d_S(\sigma) =  4 -\frac{4-d_{\rm UV}}{1+\sigma/c},
\label{CDTSD2}
\end{equation}
such that $d_{\rm UV} := d_S(\sigma \rightarrow 0) $ is the value of the spectral dimension 
in the UV limt. The parameter $c$ can be related 
to the energy scale of the dimensional reduction:  $E_* := \frac{1}{\sqrt{c}}$. 
The diffusion time $\sigma$ introduces an external parametrization of the diffusion process 
on a given space-time. Worth mentioning here is that, due to absorption of diffusion constant 
in the heat kernel equation, the dimensionality of $\sigma$ is (in the Planck units) an inverse 
energy squared.

\section{From the spectral dimension to the deformed dispersion relation}

The spectral dimension discussed in the previous section is formally defined as follows
\begin{equation}
d_S(\sigma) \equiv   -2 \frac{\partial \log P(\sigma)}{\partial \log \sigma},
\label{SpectralDim}
\end{equation} 
where
\begin{equation}
P(\sigma) =   \int d\mu   e^{\sigma \triangle_p}
\label{ReturnProbability}
\end{equation}
is the average return probability of the random walk (diffusion) process. Here, $d\mu$ is 
the invariant\footnote{With respect to the isometries of a given space-time.} measure 
in the momentum space, which we assume to be the classical one\footnote{
With this assumption we enforce that the momentum space is flat as in 
the classical case.  However, as it is hypothesized \emph{e.g.} in the context of Relative 
Locality \cite{AmelinoCamelia:2011bm}, non-vanishing curvature of the momentum 
space may be characteristic for the the quantum gravitational phenomena. This issue 
in the context of CDT will be studied elsewhere.}
$d\mu= \frac{dE d^3{\vec{p}} }{(2\pi)^4}$. $\triangle_p$ is the Laplace operator in 
momentum space.   In the case of the Euclidean 4-dimensional space 
$\triangle_p =-E^2-\vec{p}^2$. This expression might be viewed as the Wick rotated 
$(E\rightarrow iE )$ version of the analogous formula in the Minkowski space 
$\triangle_p =E^2-\vec{p}^2$, which defines dispersion relation for the free fields 
living on the space-time ($\triangle_p=0 \Rightarrow E^2=\vec{p}^2$). In this case, 
the equations (\ref{SpectralDim}) and (\ref{ReturnProbability}) predict $d_S=d$, 
where $d=4$ is the topological dimension.  

CDT predicts that the spectral dimension varies as a function of the diffusion 
time $\sigma$. The corresponding momentum Laplace operator has to be, therefore, 
deformed with respect to the classical case.  Here, we parametrize this departure 
in the following manner  
\begin{equation}
\triangle_p =-E^2- \Omega( p )^2, \label{Laplace4D}
\end{equation}
where $\Omega( p )$ is some unknown function of $p := \sqrt{\vec{p} \cdot \vec{p}}$. 

The parametrization (\ref{Laplace4D}) is of course one of many possible ones. 
In general, both the energy and momentum parts could be deformed, including 
also mixed terms. Recovering the form of the deformed momentum Laplace operator
based on a single function $d_S(\sigma)$ is, therefore, a highly ambiguous 
problem in general. This is because different forms of the deformed momentum 
Laplace operator may lead to the same (or very similar) shape of the diffusion-time 
dependence of the spectral dimension (For more detailed discussion of the ambiguity 
issue we refer to Ref. \cite{Calcagni:2013vsa}). The choice (\ref{Laplace4D}) has been 
made such that there is only a single unknown function involved, which form can
unambiguously recovered. The fact that in the applied parametrization the energy
part is not a subject of modification is supported by the CDT results. Namely, in CDT 
the time direction is imposed by time foliation and contribute classically. The same 
is, therefore, expected for energy. On the other hand, spatial slices have fractal structure, 
leading to the dimensional reduction (see Ref. \cite{Gorlich:2011ga}). In consequence, 
corrections to the momentum part of the Laplace operator are expected. Furthermore, 
the choice  (\ref{Laplace4D}) is the most conservative one and has simple physical 
interpretation because it leads to the dispersion relation in the form $E=\Omega( p )$.

\begin{figure}[ht!]
\centering
\includegraphics[width=8cm, angle = 0]{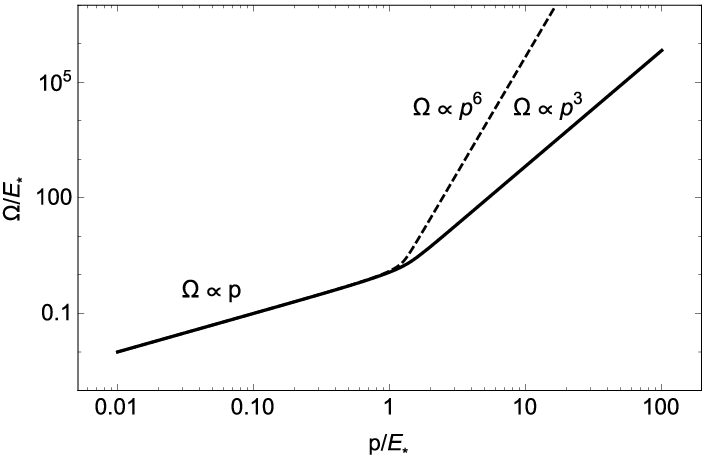}
\caption{A plot of exemplary dispersion relations predicted by dimensional reduction in 
Causal Dynamical Triangulations. The continuous line represents the  $d_{\rm UV} = 2$ case
while the dashed line corresponds to $d_{\rm UV} = 3/2$.}
\label{Disp}
\end{figure}

The task would be now to recover the form of the function  $\Omega( p )$ for the CDT  
spectral dimension (\ref{CDTSD2}). In this case, the expression for the return probability is
\begin{equation}
P(\sigma) = \frac{1}{\sqrt{4\pi \sigma}} \frac{4\pi}{(2\pi)^2} \int_{0}^{\infty}dp\ p^2
e^{-\sigma \Omega^2( p )}, 
\label{RPreduced}
\end{equation} 
where integrations over $E$ and the angular part of $d^3\vec{p}\ $ have been performed.
As it has been shown in Ref. \cite{Sotiriou:2011aa}, the relation (\ref{RPreduced}) 
can be converted into the form of the inverse Laplace transform 
\begin{equation}
p^3  =  \frac{1}{2\pi i}  \lim_{T\rightarrow \infty} \int^{\gamma+iT}_{\gamma-iT} \left(12 \pi^{5/2}  \frac{P(\sigma)}{\sqrt{\sigma}}\right) 
e^{\sigma \Omega^2( p )} d\sigma,  
\label{InverseLaplace}
\end{equation}
the $\gamma\in \mathbb{R}$ has to satisfy the condition $\forall_{\sigma_*}   \gamma> \Re(\sigma_* )$, 
where $\sigma_*$ denotes singularities of $P(\sigma)/\sqrt{\sigma}$. The $P(\sigma)$ function is recovered 
by integrating definition (\ref{SpectralDim}) with use of parametrization (\ref{CDTSD2}). We find that 
\begin{equation}
P(\sigma) = \frac{1}{16\pi^2 \sigma^{d_{\rm UV}/2}(\sigma+c)^{2-d_{\rm UV}/2}}, \label{Psigma}
\end{equation}
which is a positive definite function and, therefore, has correct probabilistic interpretation. 
By applying (\ref{Psigma}) to (\ref{InverseLaplace}) the following deformed dispersion relation
is obtained:
\begin{equation}
p^3  = \Omega^3 {_1F_1}\left(2-d_{\rm UV}/2, 5/2; -c \Omega^2\right),  
\label{DefDispRel}
\end{equation}
where ${_1F_1}(a,b;z)$ is the confluent hypergeometric function.  Equation (\ref{DefDispRel}), is an 
entangled form of the deformed dispersion relation $E=\Omega ( p )$.   The function $\Omega ( p ) $ 
for $d_{\rm UV} = 2$  and $d_{\rm UV} = 3/2$ is shown in  Fig. \ref{Disp}. 

In the IR limit $(p \rightarrow 0 )$  the dispersion relation  is approximated by 
\begin{equation}
\Omega_{\rm IR} ( p ) \approx p+\frac{E_*}{15}(4-d_{\rm UV}) \left( \frac{p}{E_*}\right)^3,    
\label{IRLim}   
\end{equation}
so the classical limit is correctly recovered. However, when passing to the high energy 
range (the UV limit) the dispersion relation undergoes deflection to the form 
\begin{align}
\Omega_{\rm UV} ( p ) \approx E_* b \left( \frac{p}{E_*}\right)^{\frac{3}{d_{\rm UV}-1}} 
\propto p^{\frac{3}{d_{\rm UV}-1}},   
\label{UVLim}       
\end{align}
where for the further convince we introduce 
\begin{equation}
b:=  \left( \frac{4 \Gamma(d_{\rm UV}/2+1/2 )}{3\sqrt{\pi}}\right)^{\frac{1}{d_{\rm UV}-1}},
\end{equation}
which is defined such that $b(d_{\rm UV}=4)=1$ and $b(d_{\rm UV}=2)=\frac{2}{3}$ 
and $b(d_{\rm UV}=3/2)\approx 0.465$. For $d_{\rm UV} = 2$ we have 
$\Omega_{\rm UV} ( p )  \propto p^3$ and  for $d_{\rm UV} = 3/2$ the $\Omega_{\rm UV} ( p )  \propto p^6$.

Both of the obtained limiting behaviors of the deformed dispersion relation will now be 
used to construct the relation of CDT to astrophysical and cosmological observations.   

\section{Astrophysics}

The dispersion relation derived in the previous section can now be used to find the group 
velocity of photons. For the particles having energy $E\ll E_*$ the IR limit approximation 
(\ref{IRLim}) may be applied, leading to the quadratic deviation    
\begin{equation}
v_{\rm gr} := \frac{\partial E}{\partial p} = \frac{\partial \Omega ( p )}{\partial p} \approx
1+\frac{3}{15}(4-d_{\rm UV})  \left(\frac{E}{E_*}\right)^2.
\label{groupvelocity}
\end{equation}
This expression predicts that the group velocity is increasing (superluminal behavior) 
or decreasing (subluminal behavior)  with the energy depending on whether $d_{\rm UV}$
is smaller or greater than four respectively.  In case of the dimensional reduction 
expected in the phase C of CDT, for a bunch of photons of different energies 
emitted simultaneously the ones with the highest energy will arrive first to the observer.  
This effect can be quantified by the arrival time difference between $E$ and  $E\rightarrow 0$
energy photons. With use of formula (\ref{groupvelocity}), the corresponding expression  is 
$\tau \simeq - L \frac{3}{15}(4-d_{\rm UV}) \left(\frac{E}{E_*}\right)^2$, where $L$ is the distance to 
a source.  Although a huge value of the energy scale $E_*$ is expected\footnote{Presumably of 
the order of the Planck energy $E_{\rm Pl} \approx 1.22 \cdot 10^{19}$GeV.}, the multiplication 
by a sufficiently large distance $L$ may bring the value of $\tau$ close to the observational 
window.  

The effect can be constrained with use of the signals from the high-energy astrophysical
sources such as  gamma ray bursts (GRB)  \cite{AmelinoCamelia:1997gz}. In particular,
the GRB 090510 remote by $L \approx 5,8 $ Mpc  may be used. Using the data
from the \emph{Fermi}-Large Area Telescope \cite{Vasileiou:2013vra} one can find that 
constraints on the energy scale of the dimensional reduction are:
\begin{align}
&E_* >   0.7 \sqrt{4-d_{\rm UV}} \cdot 10^{10} \text{GeV  at}\ (95 \% \text{CL})\ \text{for}\ d_{\rm UV} < 4, \nonumber \\
&E_* >  4.8 \sqrt{d_{\rm UV}-4}\cdot 10^{10} \text{GeV  at}\ (95 \% \text{CL})\ \text{for}\ d_{\rm UV} > 4. \nonumber
\end{align}
These constraints are definitely much stronger than any obtained from the accelerator 
physics experiments (which are reaching energies of the order of $10^{4}$ GeV). However, 
because of the quadratic (in energy) from of the effect, the constraint is distant from the 
Planck scale\footnote{If the effect had been linear in energy, the observational constraint would 
be much stronger.}  Furthermore, the obtained values of constrains are typical to 
the quadratic corrections, which are obtained in various approaches to quantum gravity  
(see Ref.  \cite{Calcagni:2016azd} for recent review).
 
On the other hand, in case when one would consider the 
high energy photons to be described by the UV part of the dispersion relation, 
the group velocity is given by 
\begin{equation}
v_{\rm gr}^{\rm UV} = \frac{\partial \Omega_{\rm UV}(p)}{\partial p} = \frac{3 b}{d_{\rm UV}-1} 
  \left(\frac{E}{b E_*}\right)^{\frac{4-d_{\rm UV}}{3}}.
\end{equation}
Such a situation is, however, very unlikely because it would require photons 
with energies $E>E_*$ to be considered and as shown previously 
this corresponds to energies above $10^{10}$ GeV.

\section{Cosmology}

Because of the quadratic nature of the IR variation of the group velocity, perhaps 
the more promising is an application of dimensional reduction to cosmology. Here, 
instead of the IR, the UV part of the dispersion relation will play a crucial role.  The 
potential relevance of the UV scaling of the modified dispersion relation in cosmology 
has been suggested in Ref. \cite{Amelino-Camelia:2013tla}. It has been shown that
dimensional reduction to $d_{\rm UV}=2$ leads to the scale invariant vacuum fluctuations 
of the primordial perturbations in the UV domain. What has not been shown is 
how the shape of the spectrum changes when we try to relate it with the one  
that is observationally relevant. Here, we show that a deviation from $d_{\rm UV}=2$ leads to 
a tilt in the power spectrum. This prediction is confronted with the observations
of the cosmic microwave background radiation (CMB).

The resulting CDT dispersion relation $E=\Omega ( p )$ can be incorporated
into the equation of modes for the cosmological perturbations.  For this 
purpose let us consider equation of modes 
\begin{equation}
\frac{d^2}{d\tau^2}f_k+\left( -\nabla^2_k - \frac{z^{''}}{z}  \right)f_k=0,  \label{ModeEq}
\end{equation}
where $\nabla^2_k$ is the momentum part of the total Laplace operator, 
which in the classical $3+1$ case is equal $\nabla^2_k=-k^2$.  
Having a given value of $k$ the corresponding physical value of
momentum can be find from the expression $p=\frac{k}{a}$, where 
$a$ is the scale factor. The value of $z$ depends on which kind of perturbations 
is considered. For the gravitational waves $z=z_T:=a$ and for the scalar 
perturbations $z=z_S:=a\frac{\dot{\phi}}{H}$. 

The task now is to phenomenologically introduce the CDT effects to Eq. \ref{ModeEq} 
by modifying the the expression for $\nabla^2_k$. This can be done by using the 
dispersion momentum part of the Laplace operator (\ref{Laplace4D}). However, 
this require a proper rescaling which will take into account relation between 
the physical momentum $p$ present in the $\Omega ( p )$ function and the 
coordinate momentum $k$ appearing while the cosmological case is considered. 
By performing the following replacement 
\begin{equation}
 -\nabla^2_k \rightarrow a^2\Omega^2(k/a),
\end{equation}
in Eq. \ref{ModeEq}, the effect of deformed dispersion relation can be introduced, such that the 
classical limit  classical limit $a^2\Omega^2(k/a) \rightarrow a^2 (k/a)^2 = k^2$ is 
recovered correctly for $p\rightarrow 0$. Worth stressing here 
is that in the above analysis we assumed that the averaged background dynamics 
is not a subject of the CDT corrections. This is satisfied in the regime where 
the homogeneous dynamics is well approximated by the classical cosmology but 
the inhomogeneities, if considered at sufficiently small spatial scales, reflect 
quantum nature of the underlying spacetime.

In consequence, in the sub-Hubble limit $\left(a^2\Omega^2(k/a) \gg  \frac{z^{''}}{z}\right)$,  the 
(Bunch-Davies) vacuum normalization of the $f_k$ mode functions leads to    
\begin{equation}
|f_k|^2=\frac{1}{2a\Omega(k/a)}.
\end{equation} 
Applying this to the definition of the scalar power spectrum we find that for $k/a \gg E_*$  
\begin{eqnarray}
\mathcal{P}_{S}(k) &&:= \frac{k^3}{2\pi^2} \frac{|f_k|^2}{z_S^2}  \nonumber \\
&&= \frac{\left( \frac{E_*}{E_{\rm Pl}}\right)^2  \left(\frac{1}{E_*} \frac{k}{a} \right)^{\frac{3(d_{\rm UV}-2)}{d_{\rm UV}-1}}}{3 \pi(1+w) b } 
\propto  k^{n_S-1},  
\label{ScalarPS}
\end{eqnarray}  
where $n_S=1+\frac{3(d_{\rm UV}-2)}{d_{\rm UV}-1}$. Furthermore, we assume that the 
universe is filled with barotropic matter: $P=w\rho$, with $w=const$.

Indeed, for $d_{\rm UV}=2$ the power spectrum (\ref{ScalarPS}) becomes scale invariant
$\mathcal{P}_{S}(k) = \frac{1}{\pi(1+w) }\left( \frac{E_*}{E_{\rm Pl}}\right)^2 =const$.    
While it is tempting to relate the obtained spectrum with that inferred from the CMB 
nearly scale-invariant spectrum of the primordial perturbations, it is not allowed to do so 
directly.  The spectrum (\ref{ScalarPS}) corresponds to the UV regime ($k/a \gg E_*$) 
which has to be suitably converted into the super-Hubble spectrum probed observationally. 
In general, this would require evolving the modes starting from the UV domain 
until crossing the Hubble horizon. This will unavoidably affect the shape of the power spectrum.   
A scenario in which knowledge of the intermediate evolution is not required corresponds to the 
case when the Hubble horizon is located in the UV domain (in the sense of the 
deformed dispersion relation). In that case \emph{freezing} of the modes occurs already 
at the level of the UV spectrum (\ref{ScalarPS}).  The freeze out  
considered here is the same as the one consider in the classical cosmology and is related 
with the fact that in the  super-Hubble limit the  Eq. \ref{ModeEq} reduces to 
$\frac{d^2}{d\tau^2}f_k - \frac{z^{''}}{z}f_k\approx 0$, such that  physical amplitude of 
perturbation i.e. $\frac{f_k}{z}$ has solution in the form $\frac{f_k}{z} = A_k +B_k \int\frac{d\tau}{z^2}$, 
where the effect of freeze out is because of non-vanishing constant of integration $A_k$.

In scenario under consideration the tensor-to-scalar ratio can be expressed as 
\begin{eqnarray}
r &&:= \frac{\mathcal{P}_{T}(k_0)}{\mathcal{P}_{S}(k_0)}  
= \frac{64 \pi G \frac{k^3}{2\pi^2} \frac{|f_k|^2}{z_T^2}}{\frac{k^3}{2\pi^2} \frac{|f_k|^2}{z_S^2}} \nonumber \\
&&=  64 \pi G  \left( \frac{z_S}{z_T} \right)^2 = 24(1+w),
\end{eqnarray}
can be directly related to the CMB data\footnote{This is possible because of growth 
of a physical distance in the expanding universe ($\lambda \propto a$).}. The best current 
observational bound at the pivot scale $k_0=0.05\ {\rm Mpc}^{-1}$ is $r<0.07$ at 95 \% confidence level
(BICEP2/\emph{Keck Array}) \cite{Array:2015xqh}, which implicates 
that $w<-0.997$ when the perturbations were formed. The universe had to, therefore, 
be very close to the de Sitter phase when the perturbations were formed. Cosmic inflation is still needed!
 
\begin{figure}[ht!]
\centering
\includegraphics[width=8cm, angle = 0]{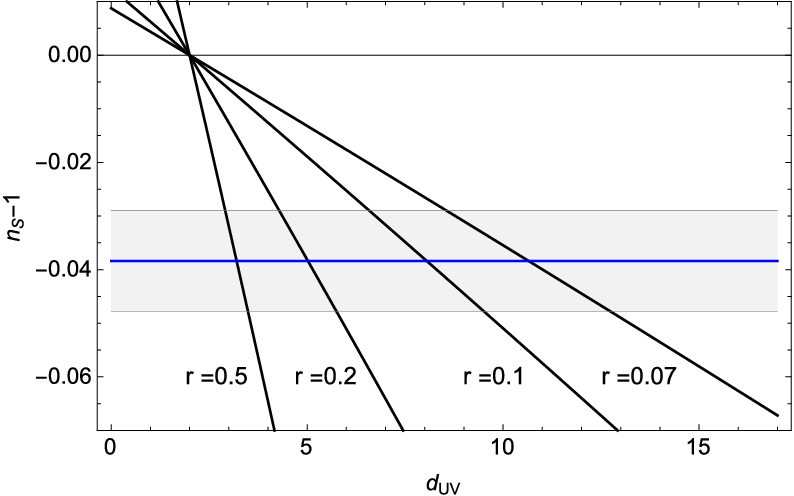}
\caption{The black lines represent function (\ref{nsrtheo}) plotted for fixed values of $r$. 
The horizontal blue line corresponds to the value $n_S-1=-0.0384$ obtained based on the 
Planck satellite measurements \cite{Ade:2013zuv}. The shadowed region represents 
uncertainty of the measurement.}
\label{nor}
\end{figure}

The value of the power spectrum accessible observationally is at the horizon-crossing. 
In our case, the horizon-crossing condition $a^2\Omega^2(k/a) = \frac{z^{''}}{z}$ 
translates into $k\simeq a E_*  \left(\frac{H}{E_* b}\right)^{\frac{d_{\rm UV}-1}{3}}$ 
$:= k_H$\footnote{Because of the deformed form of the dispersion relation, the condition differs from 
the classical expression $k\simeq a H $.}  for $H > E_*$.  Using this, the spectral index 
of scalar perturbations at the horizon-crossing is 
\begin{equation}
n_S-1 \equiv \frac{d \ln \mathcal{P}_{S}(k=k_H) }{ d\ln k} \approx \frac{3 r (d_{\rm UV}-2)}{(d_{\rm UV}-1)r-48}.
\label{nsrtheo}
\end{equation}
The scale invariance is still recovered for $d_{\rm UV}=2$.  However, according to the most up to 
date observations, the scalar power spectrum exhibits a red tilt. In particular, the \emph{Planck} satellite 
measurements provide following value of the spectral index: $n_S = 0.9616 \pm 0.0094$ \cite{Ade:2013zuv}.
 In Fig. \ref{nor} we compare predictions of the formula (\ref{nsrtheo}) with the Planck data 
for representative values of the tensor-to-scalar ratio $r$.

 Analysis of of the Fig. \ref{nor} indicates that it is impossible to explain observational 
data with the values of $d_{\rm UV}$ expected in the phase $C$ of CDT. That would 
require the value of the $r$ parameter to be greater than the observational bound $r<0.07$.
For  $d_{\rm UV}=4$ the model reduces to the classical inflation with barotropic equation 
of state with $w$ being very close to $w=-1$, which is disfavored by the observational 
data \footnote{In the language of the scalar field inflation this corresponds to the potential 
with exponential tail, which is not favored in the light of available experimental data \cite{Ade:2015lrj}.}.  
A consistency with the observational data (both the value of the spectral index
and the tensor-to-scalar ratio) can be obtained only by taking 
\begin{eqnarray}
d_{\rm UV}> 10.65.   \label{dUVconstr}  
\end{eqnarray}
However, this value is in overt contradiction with the values of $d_{\rm UV}$ expected 
in the C phase of CDT. The scenario considered can be, therefore, ruled out based on 
the CMB data. Other scenarios, such as those with $H<E_*$ at the horizon-crossing 
have to be studied separately, taking into account the evolution of modes in the intermediate 
energy range. Also, it has to be stressed that the analysis has been performed for the 
background dynamics described by the barotropic fluid. Extension of the presented 
results e.g. to the case of slow-roll is deserved. 

What we have actually shown is that the dimensional reduction does not 
provide a mechanism which can be competitive to the inflationary generation of primordial 
inhomogeneities. An attempt of using the UV-modified Bunch Davies vacuum, in particular 
corresponding to the promising $d_{\rm UV}=2$ case, did not lead to the nearly 
scale-invariant spectrum being consistent with cosmological observations. However, there 
is also a positive outcome of the performed analysis. The presented calculations indicate 
that the large values of $d_{\rm UV}$, satisfying the condition (\ref{dUVconstr}), lead the results being consistent 
with observations. Moreover, such super-diffusion type of behavior may improve fitting 
properties of the classical models which were not favored by the cosmological observations, 
as observed in the analyzed model with barotropic fluid. 

Interestingly, the increase of the spectral dimension above  $d_S=4$ has been observed in CDT 
in both, the $B$-phase and the so-called bifurcation sub-phase of the C-phase.  Further 
detailed analysis has to be performed to verify if the presented analysis can be applied to
any of these cases. Furthermore, while the super-diffusion type of behavior 
is not typically predicted within the models of Planck scale physics, it has been observed 
in Relative Locallity \cite{Arzano:2014jfa} and  in the models inspired by the Bekenstein-Hawking 
entropy-area scaling  \cite{Arzano:2013rka}. 

\section{Conclusions}

The results presented in this letter neither prove nor disprove Causal Dynamical Triangulations.  
What has been essentially shown is that phenomenology of CDT can be extracted from numerical 
studies of the spectral dimension. The predicted energy-dependence of the speed of light allowed 
us to rule out a possibility of low-energy dimensional reduction. Furthermore, by incorporating the 
predictions regarding the UV behavior of the spectral dimension to primordial cosmology, we found 
that the CDT-motivated cosmological scenarios may be observationally falsifiable. Moreover, we 
explicitly ruled out one of them with use of the observations of the CMB. 

As Thomas A. Edison once said `I have not failed. I've just found 10000 ways that won't work.'  
Quoting the sentence, we are convinced that unabated stubbornness in eliminating Planck scale 
scenarios will eventually lead to a discovery comparable only with Edison's one. 

\section*{Acknowledgements}

JM is supported by the Grant DEC-2014/13/D/ST2/01895  of the National Centre of Science.
The author would like to thank to Daniel Coumbe for his valuable comments.

\end{document}